\documentclass[aps,pra,onecolumn,notitlepage,longbibliography]{revtex4-1}
\usepackage[dvips]{graphics,graphicx}
\usepackage{amsmath}
\usepackage{epstopdf}
\usepackage{caption}
\usepackage{subcaption}

\begin{document}

\title{Chaotic and Regular Dynamics In the Three-site Bose-Hubbard
Model}
\author{A. A. Bychek$^{1}$}
\author{P. S. Muraev$^{1,2}$}
\author{D. N. Maksimov$^{1,2,3}$}
\author{A. R. Kolovsky$^{1,2}$}
\affiliation{$^1$Kirensky Institute of Physics, Federal Research Center KSC SB RAS, 660036, Krasnoyarsk, Russia\\
$^2$Siberian Federal University,  660041, Krasnoyarsk, Russia \\
$^3$Reshetnev Siberian State University of Science and Technology, 660037, Krasnoyarsk, Russia}
\date{\today}
\begin{abstract}
We analyze the energy spectrum of the three-site Bose-Hubbard
model. It is shown that this spectrum is a mixture of the regular
and irregular spectra associated with the regular and chaotic
components of the classical Bose-Hubbard model. We find relative
volumes of these components by using the pseudoclassical approach.
Substituting these values in the Berry-Robnik distribution for the
level spacing statistics we obtain good agreement with the
numerical data.
\end{abstract}

\maketitle

\section{Introduction}

The Bose-Hubbard (BH) model contains the basic physics of
interacting bosons in a lattice \cite{gersch63} with particular
interest in the context of cold Bose atoms in optical lattices
\cite{jaksch98}. The many-site BH model is known to belong to the
class of quantum nonintegrable systems whose spectral and
dynamical features are consistent with predictions of the theory
of Quantum chaos \cite{kolovsky16}. The chaotic dynamics of cold
Bose atoms in the optical lattice has been intensively studied in
recent years \cite{franzosi03, kolovsky04, mossmann06,
arwas15,kolovsky16, makarov17, khripkov19}. On the contrary, the
two-site BH model is completely integrable, i.e. can be solved
analytically \cite{franzosi00}. The cold atom realizations of the
two-site BH model are nowadays a popular playground for studying
such phenomena as Josephson oscillations and self-trapping
\cite{albiez05, gati07}.

In this work we analyze the three-site BH model which is the
simplest representative of  the nonintegrable BH Hamiltonians. On
the other hand, the three-site system retains certain features of
the integrable two-site system \cite{nemoto00,
franzosi01,franzosi03}. For example, it can show the generalized
Josephson oscillations with quasiperiodic change of the site
occupations \cite{franzosi03}. In the work we give description of
dynamical regimes of the three-site BH model and identify their
signatures in the energy spectrum. There are several ways to
distinguish regular and chaotic regimes:  Loschmidt echo
\cite{cucchietti04,goussev12}, machine learning algorithms
\cite{kharkov19}, and semiclassical (or, better to say,
pseudoclassical) methods \cite{mossmann06, graefe07}.  Here we
employ the latter approach -- we introduce the classical analogue
of the quantum three-site BH model and demonstrate that it shows a
mixture of chaotic and regular dynamics. We quantify chaos by
calculating the finite-time Lyapunov exponent  and relative
volumes of the regular and chaotic components.

\section{The system}

The Bose-Hubbard Hamiltonian reads
\begin{equation}\label{Hamiltonian}
\widehat{\cal H}=-\frac{J}{2}\sum_{l=1}^{\overline{L}} \left(
\hat{a}_{l+1}^{\dagger} \hat{a}_l + h.c. \right) +
\frac{U}{2}\sum_l^{L} \hat{a}^{\dagger}_l \hat{a}^{\dagger}_l
\hat{a}_l \hat{a}_l ,
\end{equation}
where the index $l$ labels the $l$th well of the optical
potential, $\hat{a}_l$ and $\hat{a}_l^\dagger$ are the bosonic
annihilation and creation operators,
\begin{equation}
[\hat{a}_l, \hat{a}_l^\dagger]=\hbar\delta_{l,l'},
\end{equation}
$J$ is the hopping matrix element, and $U$ the microscopic
interaction constant. Experimentally, the three-well optical
potential can be realized using different technics \cite{esteve08}
where the hopping energy and the particle interaction can be
controlled separately.

Depending on the lattice geometry of the summation limit in the
hopping term $\overline{L}$ can take two different values. If the
potential wells are arranged along a straight line the system is a
linear oligomer (LO) where the hopping terminated at the first
$l=1$ and the last $l=L$ sites on the line. Then we have
\begin{equation}\label{LO}
\overline{L}_{LO}=L-1.
\end{equation}
Another geometry is a circular oligomer (CO) where we additionally
have the hopping between the first and last sites that leads to
the periodical boundary condition $\hat{a}_{L+1}=\hat{a}_1$. Then
\begin{equation}\label{CO}
\overline{L}_{CO}=L.
\end{equation}
The periodic boundary condition also implies conservation of the
total quasimomentum. This can be proved by rewriting the
Hamiltonian (\ref{Hamiltonian}) in terms of the operators
$\hat{b}_k$ and $\hat{b}_k^\dagger$,
\begin{equation}\label{Bloch_basis}
\hat{b}_k=\frac{1}{\sqrt{L}}\sum_l \exp(i 2\pi k l/L) \hat{a}_l.
\end{equation}
which annihilate and create a particle in the Bloch state with the
quasimomentum $\kappa = 2\pi k/L$. We obtain
\begin{equation}\label{Hamiltonian_Bloch_basis}
\widehat{\cal H}=-J\sum_{k=-1,0,1} \cos{\left(\frac{2\pi
k}{3}\right)} \hat{b}_k^\dagger \hat{b}_k +
\frac{U}{6}\sum_{k_1,k_2,k_3,k_4} \hat{b}_{k_1}^\dagger
\hat{b}_{k_2}^\dagger \hat{b}_{k_3} \hat{b}_{k_4} \tilde{\delta}
(k_1+k_2-k_3-k_4),
\end{equation}
where the presence of the $\delta-$function in the interaction
term insures that the total quasimomentum is conserved. The
Hilbert space of (\ref{Hamiltonian_Bloch_basis}) is spanned by the
quasimomentum Fock states $| n_{-1}, n_0, n_{+1} \rangle \ $,
where $\Sigma_k n_k =N $ is the total number of particles.
\begin{figure}[h]
\center
\includegraphics[width=0.8\textwidth]{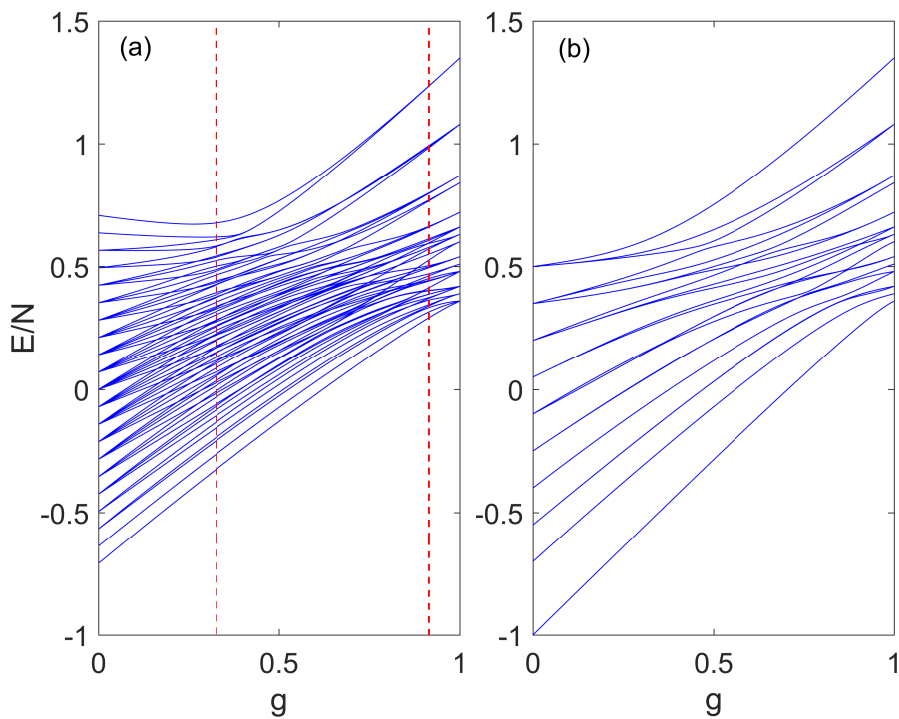}
\caption{Energy spectrum of $N=10$ bosons as a function of the
macroscopic interaction constant $g=UN/3$ and the hopping matrix
element $J=1-g$. Left panel: the whole spectrum of the linear
oligomer (LO). Right panel: the spectrum of the circular oligomer
(CO) for the independent subset of states with zero total
quasimomentum.} \label{fig1}
\end{figure}

\begin{figure}[h!]
\center
\includegraphics[width=0.6\textwidth]{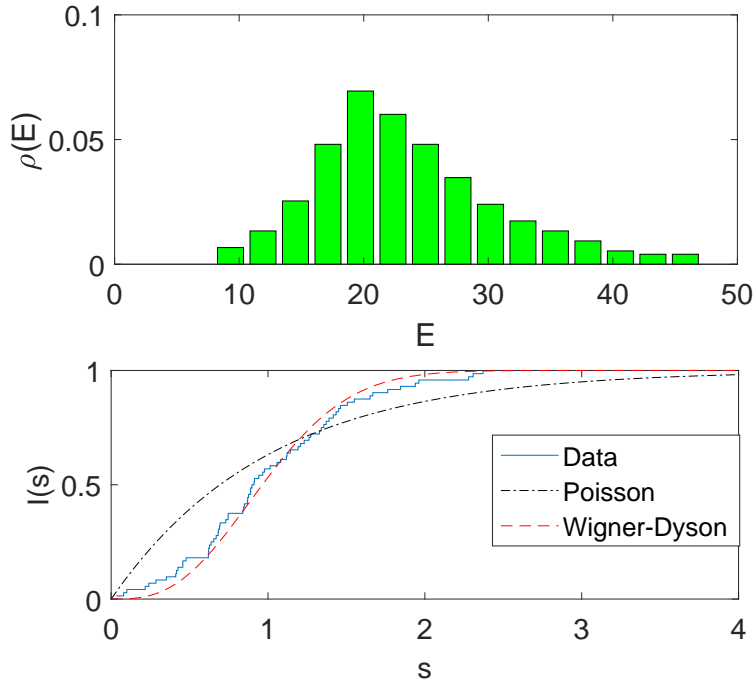}
\caption{Upper panel: The density of states $\rho$ of the
three-site Bose-Hubbard model. Lower panel: Integrated
distribution $I(s)=\int_{0}^{s} ds' P(s')$ for the Poisson level
spacing statistics $P_P(s)$ (dash-dotted black curve), the
Wigner-Dyson statistics $P_{WD} (s)$ (dashed red curve), and
numerical data $P (s)$ (blue) for the central part of the energy
spectrum. Parameters are $N=30$,  $g=0.8$, and $J=1-g$.}
\label{fig2}
\end{figure}

Next we discuss the energy spectrum of the system. Fig. \ref{fig1}
shows the energy spectrum for LO (left panel) and CO (right panel)
as a function of the macroscopic interaction constant $g=UN/3$,
where we simultaneously set the hopping matrix element to $J=1-g$.
This allows us to consider the both  cases of weak and strong
coupling/interaction -- the case of $g = 0$ corresponds to the
system of noninteracting bosons whereas in the case of $g = 1$ the
interwell tunnelling is completely suppressed. In these limits the
BH model is integrable and its energy spectrum can be found
analytically. However, within the intermediate range of $g$ it is
highly irregular and the energy levels exhibit avoided crossings
as they approach each other, see the area between two dashed red
lines in Fig. \ref{fig1}(a). In this parameter region the system
is nonintegrable. Following the standard procedure we calculate
the normalized distances between the nearest levels
$s=(E_{n+1}-E_n) \rho (E_n)$ where $\rho (E)$ is the density of
states, see Fig. \ref{fig2}(a). In Fig. \ref{fig2}(b) we show the
integrated level spacing distribution $I(s)=\int_{0}^{s} ds'
P(s')$ for the central energy region comprising 70 percent of the
states (solid line) and compare it with the Poisson distribution
(dash-dotted line),
\begin{equation}\label{Poisson}
P_P (s)=exp(-s),
\end{equation}
and the Wigner-Dyson distribution (dashed line),
\begin{equation}\label{Wigner-Dyson}
P_{WD} (s)=\frac{\pi}{2} s \, exp(- \frac{\pi}{4} s^2).
\end{equation}
It is seen that the level spacing statistics is close to the
Wigner-Dyson distribution which is a hallmark of Quantum chaos.

We stress that we get a reasonable agreement with the Wigner-Dyson
statistics only because we neglect 30 percent of the energy levels
which are presumably not chaotic. A more accurate description of
the spectrum  is given by the  Berry-Robnic distribution,
\begin{equation}\label{Berry-Robnik}
P_{BR} (s)= \left[\nu_r^2 erfc(\frac{\sqrt{\pi}}{2} \nu_c s)+(2
\nu_r \nu_c + \frac{\pi}{2} \nu_c^2 s)exp(-\frac{\pi}{4}\nu_c^2
s^2)\right] exp(-\nu_r s),
\end{equation}
which includes  the relative size of the regular $(\nu_r)$ and
chaotic $(\nu_c)$ components as the fitting parameters. In the
next section we obtain these fitting parameters from the first
principles by using the pseudoclassical approach.

\section{Regular and chaotic dynamics}

Pseudoclassical approach borrows its ideas from the semiclassical
method in single-particle quantum mechanics to address the
spectral and dynamical properties of the system of $N$ interacting
bosons with $1/N$ playing the role of Planck's
constant\cite{mossmann06,graefe07,zibold10,bychek18}.  In this
approach the operators are substituted by their Weyl images which
gives
\begin{equation}\label{C_functions}
\frac{\hat{a}_l}{\sqrt{N}} \rightarrow a_l ,\quad
\frac{\hat{a}_l^\dagger}{\sqrt{N}}\rightarrow a_l^* ,
\end{equation}
and
\begin{equation}\label{classical_Hamiltonian}
\frac{\widehat{H}}{N} \rightarrow H=-\frac{J}{2}
\sum_{l=1}^{\bar{L}}\left({a}_{l+1}^* {a}_l + c.c. \right)
+\frac{g}{2}\sum_{l=1}^{L} |a_l|^4
\end{equation}
where $g=UN/L$ is the macroscopic interaction constant. In the
semiclassical limit  $N \rightarrow \infty$ and $U=g/N \rightarrow
0$ this approach is equivalent to the mean-field approximation.
The main advantage of the pseudoclassical approach above the
mean-field approximation is that it can treat the case of finite
$N$ as well. The validity of this approach was discussed, for
example, in Ref. \cite{graefe07} where it was demonstrated that it
works well till $N\sim 10$. In what follows, however, we shall
assume the limit $N\rightarrow\infty$ where the  Hamiltonian
(\ref{classical_Hamiltonian}) generates  the Hamilton equations of
motion

\begin{equation}\label{3.1}
i \frac{da_l}{dt}= \frac{\partial H}{\partial
a_l^*}=-\frac{J}{2}\left(a_{l-1}+a_{l+1} \right)+g |a_l|^2 a_l,
\hspace*{0.5cm} i \frac{da_l^*}{dt}= -\frac{\partial H}{\partial
a_l},
\end{equation}
which are known as the discrete nonlinear Schr{\"o}dinger equation
(DNLSE) \cite{smerzi03}. The solution  $a_l (t)$ is the classical
trajectory and, since ${|a_1 (t)|^2 + |a_2 (t)|^2+|a_3 (t)|^2=1}$,
it is bounded to the $S^5$ sphere  in the 6-dimensional phase
space.

\begin{figure}

        \includegraphics[width=\linewidth]{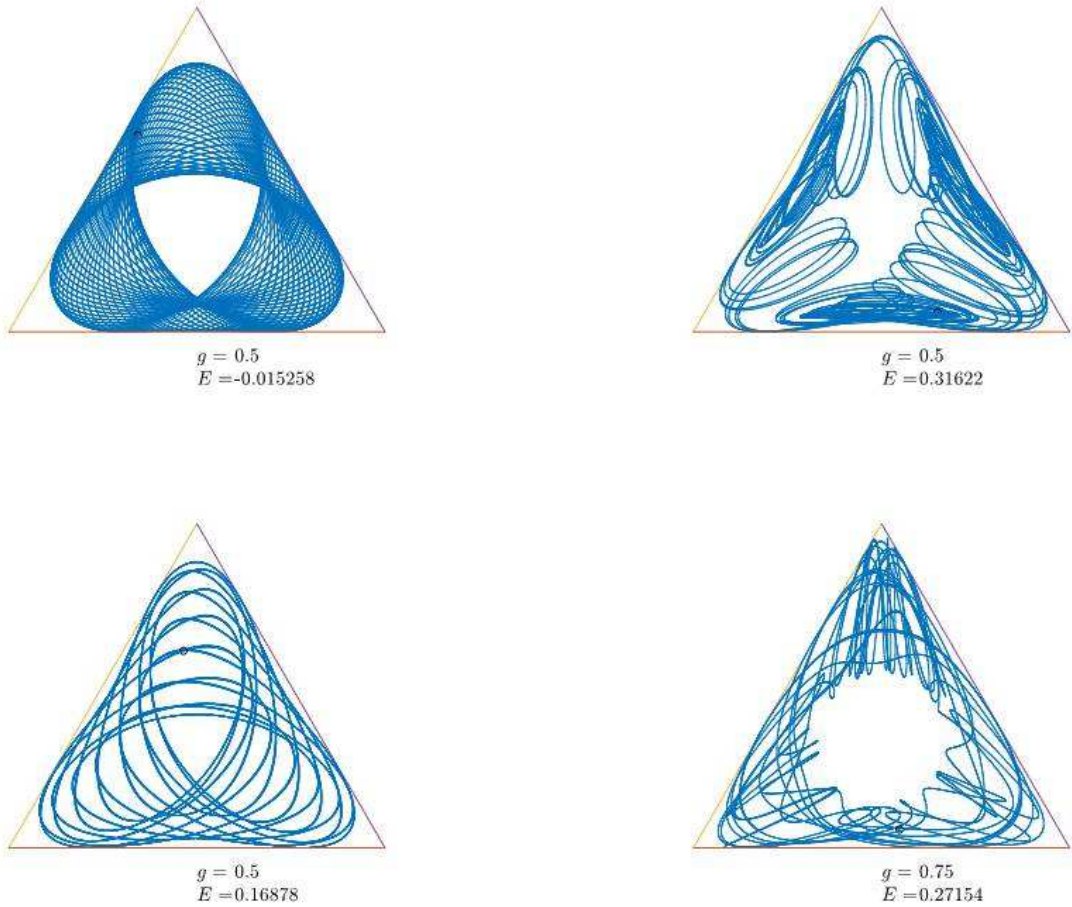}

\caption{Examples of regular (a) and chaotic (b) trajectories.
Regular dynamics correspond to the Josephson oscillations of the
site occupations.} \label{fig3}
\end{figure}

\begin{figure}[h!]
\center
\includegraphics[width=0.6\textwidth]{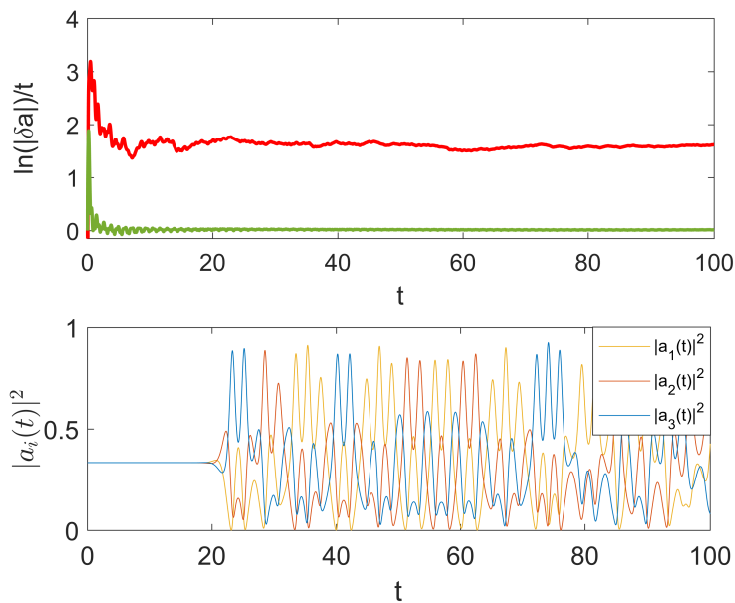}
\caption{Upper panel: Finite-time Lyapunov exponent $\lambda(t)$
for regular ($\lambda \simeq 0|_{t=t_{end}}$) and chaotic
($\lambda > 0|_{t=t_{end}}$) trajectories. Lower panel: The
phenomenon of the dynamical or modulation instability for the
Bloch wave with nonzero total quasimomentum due to chaoticity.}
\label{fig4}
\end{figure}

We numerically solve Hamilton's equations of motion (\ref{3.1})
for the ensemble of initial conditions uniformly distributed over
the whole phase space, i.e., over the surface of the sphere $S^5$.
As an example, Fig. \ref{fig3} samples regular and chaotic
trajectories from this ensemble. We distinguish between regular
and chaotic trajectories by  calculating the finite-time Lyapunov
exponent defined according to the following equation
\cite{kolovsky09,sandor04,prants02}
\begin{equation}\label{Lyapunov_exp}
\lambda (t)=\frac{ |\delta \mathbf{a} (t)| }{ |\delta
\mathbf{a}_0| }/t.
\end{equation}
Here   $\delta \mathbf{a} = (\delta a_1, \delta a_2, \delta a_3,
\delta a_1^*, \delta a_2^*, \delta a_3^*)^T$ is the deviation from
a given trajectory $\mathbf{a}(t)$ which obeys the linearized
equation of motion
\begin{equation}\label{3.2}
 i\frac{d}{dt}\delta \mathbf{a} = M[\mathbf{a}(t)] \delta \mathbf{a},
\end{equation}
with $M[\mathbf{a}(t)]$ being  $2L \times 2L$ matrix of the
following structure:
\begin{equation}\label{eq_2}
  M[\mathbf{a}(t)] = \left(\begin{array}{cc}
                    A+gB & gC \\
                    -gC^* & -(A+gB)^*
                  \end{array}\right),
\end{equation}

\begin{equation}\label{eq_3}
  A_{l, m} = -\frac{J}{2}(\delta_{l+1, m}+\delta_{l-1, m}),
\end{equation}

\begin{align}\label{eq_4}
  B_{l, m} = (2|a_l(t)|^2-\frac{E}{g})\delta_{l,m} & \qquad C_{l, m} = a_l^2(t)\delta_{l,m}
\end{align}

\begin{equation}\label{eq_5}
  E = \frac{1}{3}g - J\cos(\frac{2\pi}{3}k)
\end{equation}
The finite-time Lyapunov exponent shows the temporary evolution of
the separation between two close initial conditions and for
sufficiently long computational times converges to the celebrated
Lyapunov exponent. Namely, it approaches zero for regular
trajectories ($\lambda \simeq 0$) while it is always well above
zero for chaotic trajectories ($\lambda \geq 0$),  see Fig.
\ref{fig4}(a).

\begin{figure}[h!]
\center
\includegraphics[width=0.9\textwidth]{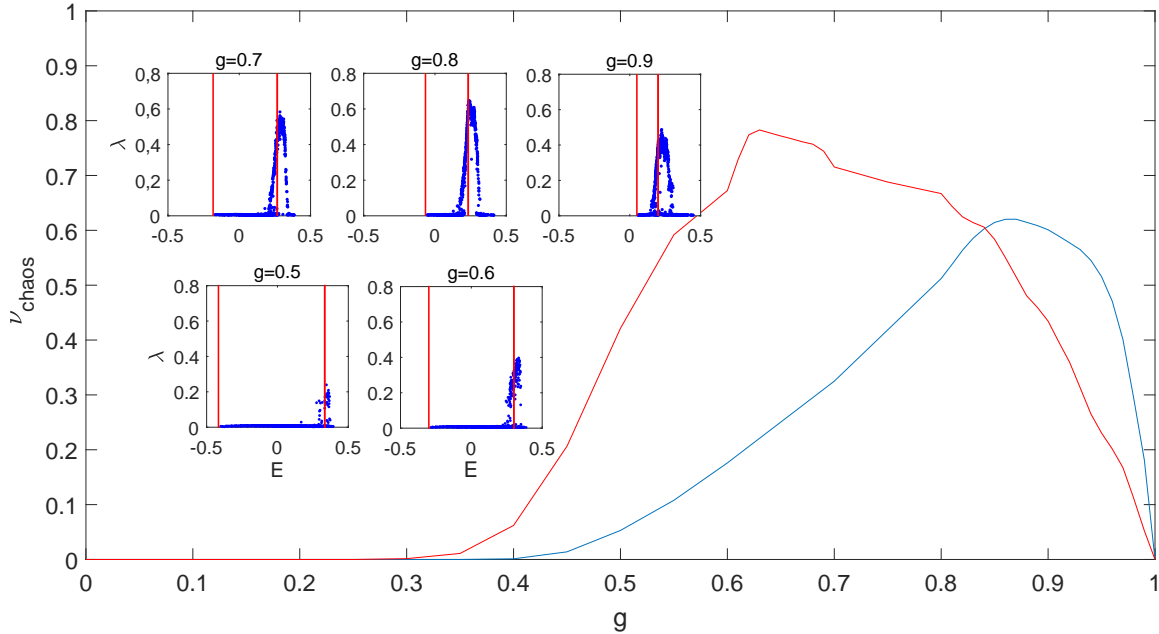}
\caption{The volume of the chaotic component as a function of the
macroscopic interaction constant $g$ for thelinear oligomer (red
curve) and circular oligomer (blue curve). Results are based on
the numerical analysis of the ensemble of $500$ trajectories with
initial conditions uniformly distributed over the whole phase
space. Insets show the Lyapunov exponent for each trajectory from
the uniform ensemble as a function of the trajectory energy given
by Eq.~(\ref{classical_Hamiltonian}). Vertical lines mark energies
of the periodic trajectories (\ref{3.3}).} \label{fig5}
\end{figure}

Next, we calculate the exponent $\lambda$ for all trajectories
from the uniform ensemble of initial conditions. The insets in
Fig. \ref{fig5} show $\lambda$ as a function of the trajectory
energy $E$ (which is obviously a conserved quantity). Additional
vertical lines mark the energies of the nonlinear Bloch waves,
\begin{equation}\label{3.3}
a_l (t)=\frac{1}{\sqrt{L}} exp [i\kappa l + i J cos(\kappa) t -i g
t], \quad \kappa=2 \pi k/L,
\end{equation}
which are stable ($|\kappa|< \pi/2$) or unstable
($|\kappa|>\pi/2$) periodic trajectories of the system, see Fig.
\ref{fig4}(b). We count the number of regular and chaotic
trajectories by introducing some $\lambda_{cr}\ll 1$ which we set
in our simulations to $\lambda_{cr}=0.01$. Then all trajectories
with finite-time Lyapunov exponent $\lambda< \lambda_{cr}$ are
treated as regular. Following this idea we find volumes of the
regular and chaotic components  as the relative number of regular
and chaotic trajectories. The results are shown in the main panel
in Fig. \ref{fig5} where the blue curve refers to the case of
circular olligomer and the red curve to the linear oligomer.
%
\begin{figure}[h!]
\center
\includegraphics[width=0.8\textwidth]{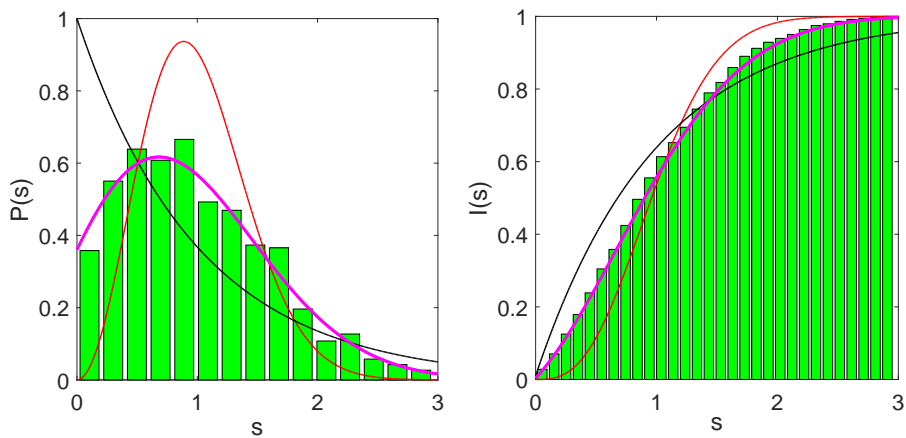}
\caption{The level spacing distribution (left panel) and
integrated level spacing distribution (right panel) of the quantum
energy spectrum in the comparison with the Berry-Robnic
distribution (magenta line) with $v_r$ and $v_c$ extracted from
results of the pseudoclassical analysis. Additionally, the red and
black lines show the Wigner-Dyson and Poisson distribution.
Parameters: linear oligomer, $v_c=0.8, v_r=0.2, N=50, g=0.8,
{J=1-g}$. } \label{fig6}
\end{figure}
%

\section{Results and conclusions}

Now we are prepared to discuss the statistical properties of the
whole energy spectrum of the three-site BH system. The histograms
in Fig. \ref{fig6} show the level spacing distribution and
integrated level spacing distribution as compared to the
Berry-Robnic distribution (\ref{Berry-Robnik}) which is depicted
by the magenta line. In the distribution (\ref{Berry-Robnik}) we
use the values of the parameters $v_c$ and $v_r$ obtained in the
previous section.  A nice agreement is noticed. This agreement
proves that the three-site BH model is a genuine mixed system
where the regular spectrum coexists with the irregular spectrum.
Furthermore, the results presented in the insets in Fig.
\ref{fig5} undoubtedly tell us which part of the energy spectrum
is associated with the chaotic dynamics and, hence, is irregular.

\begin{acknowledgments}
This work has been supported through Russian Science Foundation
Grant N19-12-00167.
\end{acknowledgments}

\nocite{*}

\bibliography{draft}

\end{document}